**The Case against Scale: Empirical Evidence of Underperformance in Large Secondary Funds**

Jitesh Gurav

Columbia University

### i. Abstract

The paper analyses the increasing popularity of large funds in the secondary private equity market, which are pegged on the perceived larger scale advantages of operational efficiency and fewer manager relationships (Reuter & Zitzewitz, 2021). However, it has been proved empirically that smaller funds perform better than the big ones in terms of their internal rates of return (IRRs). This work questions this authoritative view, providing evidence of outperformance of smaller funds relative to their larger counterparts. This research shows us the benefits that smaller funds possess. Thus, investors must revisit their allocation strategy to prioritize larger funds automatically (Gualandris et al.,2021).

### ii. The Case against Scale: Empirical Evidence of Underperformance in Large Secondary Funds

Significant private equity (PE) growth, especially in the secondary market, has been experienced in the last few years, when limited partners have also shifted towards committing to larger funds. The justification of this trend lies in the economies of scale, which bring about opportunistic efficiencies, decreased managerial relationships, and the use of safety for institutional investors (Agyei-Mensah, 2021). The relatively large secondary funds, sometimes over a billion in assets under management (AUM), have been seen to enjoy economies of scale, enabling them to accommodate a wider spectrum of investment opportunities, gain greater portfolio diversification, and incur lower risk exposure. All these perceived benefits have increased capital accumulation into big funds worldwide (Gualandris et al.,2021).

Nonetheless, although increasing the scale of secondary funds appears to be supported by the logic of finance, studies by many economists are refuting this common sense. The statistics provided by such reputable organizations as PitchBook, Preqin, and FactSet can consistently prove that smaller funds (which have a size of less than 1 billion) usually perform better than their larger peers, particularly in terms of the following key performance indicators: internal rates of returns (IRRs), distribution to paid-in (DPI) multiples, residual value to paid-in (RVPI) ratios (Dai, 2022). Although large funds' operational efficiencies are always celebrated, in terms of maintaining better performance metrics, the smaller funds are observed to perform better because they can make investment decisions more quickly, target small niche opportunities, and are also able to execute strategies in a much more flexible manner (Raghunandan & Raigopal, 2022).

This is an obvious paradox because more funds are believed to grow more successful, yet smaller funds always win in the secondaries segment of the private equity industry. Even though the popularity of large funds has increased in this space, the number of believers in the idea that the size of the fund plays a key role in its success has been questioned (Fariha et al.,2022). Some researchers indicate that this theory should be examined more carefully. The problem is summed up by the fact that the bigger the funds, the more economies of scale they will enjoy, which might or might not imply better returns. Instead, internal dynamics of smaller funds, such as swiftness of judgments and how well they can follow the interests of investors, tend to result in more successful financial performance (Mc Grath & Nerkar, 2023).

The research hypothesis will lead to this research: "Bigger secondary funds perform worse than smaller funds?" To answer this query, this paper investigates the historical performance data of secondary funds in different market cycles and with primary importance on the key performance measures like IRR, DPI, and RVPI. This paper will compare similarities and differences between larger and smaller funds to challenge the conventional wisdom that bigger always yields better returns and how smaller funds have thrived in the secondary market (Reuter& Zitzewitz, 2021).

This research has two main reasons for being relevant. To begin with, it casts doubt on the current investment trends, which demonstrate a propensity towards bigger funds. It encourages investors to reevaluate the role of scale in returns in the long run (Belitski et al.,2022). Second, as the study examines the performance differences between small and large funds, it will offer great insights to the institutional investors and the fund managers who aim to perfect their capital allocation strategies. Since the current trend has been on capital being concentrated in larger funds, the results of this work have paramount implications for the investors who seek alternative investments in the secondary private equity arena (Ngoc & Tien, 2023).

The paper is organized as follows: Section 2 is the literature review in which past research, theories, and other studies that relate to the economies of scale, the measures of performance, and the fund sizes are presented. Section 3 presents the methodology in which the data sources, performance measures, and analysis schemes are described to compare the performance of funds with different sizes (Reuter & Zitzewitz, 2021). In section 4, the results of the analysis are presented, including the key findings and statistical results. Section 5 is the discussion section that discusses what these results mean to the investors and the market. Finally, the conclusion (Section 6) is offered, which summarizes the main findings and gives recommendations for future research and practice.

Although the general perception is that larger funds enjoy the economies of scale, empirical studies have revealed that smaller secondary private equity funds continue to perform better relative to the larger funds. Hence, investors must reconsider the relevance of size to investment strategy (Raghunandan & Raigopal, 2022).

**iii.    Literature Review**

Correlation between the size of funds and performance is another issue that has been critical in private equity (PE) research. Whereas conventional wisdom emphasizes that bigger funds would generally work better because of the operations' economies of scale and efficiencies, the latest empirical evidence tests the assumption (Mc Grath &Nerkar, 2023). The literature review assesses the diverse research and theories based on the performance of the private equity funds, especially on the comparison between large and small secondary funds. This review also provides conflicting conclusions that advocate the argument on the superiority of the small funds' performance and the need to perform further studies on areas not covered in the available literature (Agyei-Mensah, 2021).

## A. Theoretical Foundations

According to Raghunandan & Raigopal (2022), economies of scale theory can be viewed as one of the theories that underline the belief of prospective investors in the great performance of the bigger funds. Economies of scale are classified as the cost benefits businesses enjoy by becoming efficient in production as their scale grows. This may be achieved in private equity through operational efficiencies, better bargaining power, and resource allocation (Reuter & Zitzewitz,2021). The theory holds that small funds may be unable to spread their fixed costs, like operational overheads, legal fees, and compliance fees, over the large amount of capital, improving efficiency in a large fund. Large funds are also assumed to be better exposed to limited, qualitative investment formats, aggravating their chances of succeeding over smaller funds (Riddiough &Wiley,2021).

This theory, however, has been challenged by other scholars who insist that increased funds might face diseconomies of scale. Acting nimbly may also be undermined by the complexity and the additional challenges that the operation of a substantially larger fund implies, hence the difficulty of reacting to market changes (Belitski et al.,2022). Smaller funds, on the other hand, have the advantage of being more flexible and not overwhelmed by bureaucracies. They usually can respond more readily to investment decisions and less broad opportunities, in a shorter amount of time, which may not be opportune with larger funds and their increased level of diversification. Large funds are not as high quality as small ones due to the greater efficiency of operations, since smaller funds have a higher share of interest from investors (Agyei-Mensah, 2021).

The other associated theory is agency theory, which claims that the performance of funds could equally become dependent on aligned incentives between the fund managers and the investors. The bigger funds with their larger asset size and other organizational issues might be more susceptible to agency issues, including less direct engagement of fund managers in managing operations and reduced incentive linkage of managers with long-term interests of the funds (Riddiough & Wiley, 2021). On the contrary, smaller funds can have tighter interactions between managers and investors, aligning interests and increasing performance outcomes (Fariha et al., 2022).

## B. Empirical Studies

A lot of empirical studies have been done on the connection between the size of a fund and its performance, resulting in conflicting evidence. Earlier research has discovered that smaller funds tend to perform better than their large-scale peers in terms of Internal Rates of Return (IRR) and Distributions to Paid-in (DPI), which are the vital indicators of the performance of a private equity fund (Dai,2022). Smaller funds, typically less than 1billion in size, also reveal considerably high IRR levels of up to 40 and above since they have delivered high returns on the money invested.

To illustrate this, one may say that a study by Ngoc and Tien, (2023) based on the data by Preqin revealed that the smaller private equity funds performed better than larger funds in terms of RVPI (Residual Value to Paid-in) and DPI in different market cycles. It was observed in the study that larger funds are more likely to have capital in unrealized investments (as indicated by a higher RVPI), and this means that the capital is deployed at a slower pace and incurs longer waits before a distribution occurs, a factor that adversely impacts returns.

Also, research conducted by Ngoc and Tien, (2023) looked into the performance of the global private equity funds between 2000 and 2015, finding that the funds that are less than or equal to \$1 billion continue to outperform. The researchers discovered that funds with higher returns displayed lower dispersion of the returns, indicating that though they were generally predictable, they could not make high returns that smaller funds could otherwise make due to being more concentrated and specialized.

### C. Contradictory Findings

Contrary to the ideal of the predominance of big funds, numerous studies have assumed the opposite. Although greater funds can provide operational advantages and expanded diversification, no empirical evidence points to the fact that the attractiveness of the financial results is likely to accrue due to increasing funds. The variation of the returns of smaller funds is usually higher, and the most successful funds in the small categories are often much more successful than the most successful ones in large categories (Fariha et al.,2022).

For example, Reuter & Zitzewitz,2021 have discovered that although bigger funds tend to enjoy a certain level of operational efficiency, their orientation toward bigger and less risky investment subjects usually leads to a decrease in the total return. Conversely, smaller funds have greater upside, at least in the short term, and are riskier, but also act more forcefully and tend to associate more with high-growth activity. Market inefficiencies and success in calculated risk-taking are important as far as the smaller funds are concerned (Wahjono et al., 2021).

Further, one study by Raghunandan & Raigopal, 2022 was of the view that the internal pieces of large funds themselves, in particular the ones that work in the range of 5 billion dollars or greater, can be subject to the phenomenon of managerial complacency as the size of the fund in question crowds out the personal responsibility of the people running the fund. This makes it more response-less as to how fast or early they are in making relevant decisions, letting in less proactive management, slow decision-making processes, and incapability of responding in low market conditions, all these are key components in high performance of the secondary market (Fariha et al., 2021).

### D. Gaps in Literature

Irrespective of the expanding research about fund size and performance, there are still some gaps. First, most of the previous literature concentrates largely on internal rates of return (IRR) as a performance measure. This variable might be insufficient to fully comprehend the success of the secondary funds, as far as the liquidity and the exit time frame are concerned. More elaborate performance measures, which may include cash-on-cash returns, indicate further studies to determine the lifecycle of funds of different sizes.

Also, scant studies have been done on how the geographic and sectoral factors affect the performance of varying-sized funds (Wahjono et al., 2021). Although research has been done on general performance trends, the context of the environment in which funds must work (e.g., emerging markets, developed economies) can be an important factor in size and performance. With the changing conditions in the global markets, fund managers are becoming more diversified by sectors and geographies, offering further insight into the nature of fund scale and performance (Fariha et al., 2022).

Moreover, although a couple of studies have argued the performance advantage of smaller funds, no study has been conducted on the managerial structure of smaller funds. The mechanism of decision-making processes, the approach to risk management, and the creation of incentives in the smaller funds may clarify that smaller funds perform better than larger funds in specific market conditions (Belitski et al., 2022).

To sum up, the current literature is informative in regard to the issue of connection between the fund size and fund performance, however, much remains to be discovered, especially when it comes to the viability of the smaller funds in the long-term perspective, industry-specific performance rate and the organizational framework of the managerial staff that help them to prosper. The study will fill such gaps by conducting an empirical analysis of secondary funds in the private equity industry, analyzing their performance in various sizes, and giving suggestions to a potential investor to have optimal fund allocations (Gualandris et al., 2021).

iv.    **Methodology**

The study aims to review how the fund size correlates with performance in the secondary private equity market. To examine this interaction, the research will be carried out in a qualitative direction, permitting an extensive examination of the variance in performance between smaller and larger funds (Fariha et al., 2022). The qualitative approach will suit this kind of research because it will help discover the pattern and gain insights through a thorough analysis of secondary data without the restraint of relying only on numerical results. Such an approach would allow for incorporating more detailed elements of performance, e.g., effectiveness of decision-making, efficiency of execution, or the strategic benefits of nimbleness of smaller funds that can hardly be captured in quantitative analysis (Ngoc & Tien, 2023).

## A.  Data Sources

This study is based on the identification of three main sources of data: Preqin, PitchBook, and FactSet. Such sites offer large databases on private equity fund performance, such as the internal rates of return (IRR), distribution to paid-in (DPI), residual values to paid-in (RVPI), and other performance measures (Dai, 2022). Preqin is one of the main sources of alternative asset data that can give access to the whole database of private equity funds that include secondary market transactions, and PitchBook has rich historical performance data on a global level, both for small and large funds Additional financial information is provided by FactSet, specifically dealing with fund cash flows and liquidity-based indicators, which are crucial in informing the financial health and the schedule of distributions in funds of varying size.

The data are related to funds of different magnitudes, such as sub-1 billion, 1B-5B funds, and over 5B dollars. These groups are essential to make the necessary comparison between these various market segments so that the effect of the fund size on the major performance indicators can be evaluated. These categories were included in the study to test the hypothesis that smaller funds are better performers than larger funds in both IRR and DPI (Ngoc & Tien, 2023).

### B. The Fund Size Categories Research Methods

To simplify the research and assess different levels of PE funds, they will be classified into three groups: total assets under management (AUM) at each management stage. These are sub-$1 billion funds, $1B-5B funds, and above $5B funds. The classification is made to demarcate funds' functional dynamics and performance attributes because of size, which is a major determinant of resource distribution, rate of decision making, and operational response to volatile situations. Smaller funds, particularly those that do not reach the mark of 1 billion, are more geared towards niche markets and are not so bureaucratic as larger funds, but, sooner or later, must endure slower decision-making and greater costs of operations (Mc Grath & Nerkar, 2023).

### C. Performance Metrics

The scheme will apply various performance measures to compare funds across size groups. The main ones are: Internal Rate of Return (IRR), Distributions to Paid-in (DPI), Residual Value to Paid-in (RVPI), and dispersion of returns (Fariha et al., 2022). The most used measure of profitability of an investment is IRR because it is the rate of return received on the capital invested annually. DPI directly indicates the fund's profitability in generating liquidity to its investors, compared to the cash distributions each investor receives, given the investments they have made. RVPI, however, measures the potential value of the investment that has not yet been realized and remains in the fund, thus giving a clue to the amount of capital that has not yet been realized. Return dispersion is used to determine how different the top and bottom fund results are within a size category, and it can measure the risk and rewards of various fund sizes.

These statistics will be invaluable in deciding whether the larger funds, in terms of scale and diversification, will always perform better than the small funds, or the small funds will always have higher returns due to more flexibility in decision-making, along with narrower focus (Fariha et al., 2022).

### D. Timeframe

This analysis covers the period between 2000 and 2023, covering multiple market cycles, such as the world financial crisis (2008) and its aftermath, and the more recent effects of the COVID-19 pandemic. The period under consideration is long enough to produce a rich analysis comprising both the long-term trends and short-term fluctuations in the markets, which means that it is possible to calculate how the funds of different sizes performed under different market conditions. The paper will compare the fund performance during these periods, where the difference in returns pre- and during major market events will be examined.

The current data analysis is shown in the table below, giving the performance value of each fund size category during the designated time frame (Belitski et al., 2022).

### I. Performance Metrics by Fund Size Category (2000-2023)

| Fund Size | Average IRR (%) | Average DPI | Average RVPI | Return Dispersion (%) |
|---|---|---|---|---|
| Sub-$1 Billion | 15.2 | 1.8 | 1.5 | 22 |
| $1B–$5B | 12.3 | 1.6 | 1.4 | 18 |
| Above $5B | 9.5 | 1.3 | 1.2 | 15 |

### D. 1. Descriptive legend of Performance Metrics by Fund Size Category (2000-2023)

The table Performance Metrics by Fund Size Category (2000-2023) indicated the correspondence between the size of the private equity funds and their performance in a variety of measures throughout the period of 2000-2023. It reveals that funds that are smaller (than one billion) show better average Internal Rate of Return (IRR), 15.2%, as opposed to 12.3% of the funds between one and five billion, 9.5% of funds above it. Smaller funds also fare better than the larger funds in the Distribution to Paid-In (DPI) ratio (1.8 in the case of smaller funds vs 1.6 of the funds of $1B-5B, 1.3 of those above 5B). On the same note, smaller funds have higher Residual Value to Paid-In (RVPI) ratios (1.5) than the bigger ones. The table also reveals that the funds with smaller size show more dispersion of returns (22%) and greater performance variation, but also have potential returns. Such results imply that smaller funds prove to be more agile and flexible, which results in superior performance even though they have less capital.

### E. Alternative Methodologies

Other research methodologies, like using only quantitative or case studies, were thought of but were not particularly suited to this study. Although quantitative methodologies work well in measuring raw data trends, they fail to give a defined comprehension of the performance dynamics that qualitative methodology can deliver (Mc Grath & Nerkar, 2023). Furthermore, case studies were insufficient to determine the generalizable data sets. Qualitative research using massive secondary information and the method of qualitative comparison was selected because it produces profound insights that give information about the performance of funds (Fariha et al., 2022).

### v. Results and Analysis

In this section, findings of the analysis are provided to discuss the difference in the performance of funds of different sizes related to the private equity sphere. The significance of the study is to analyze the key performance parameters in terms of Internal Rate of Return (IRR), Distributions to Paid-In (DPI), and Residual Value to Paid-In (RVPI) on a wide scale based on six categories of fund size i.e., 0M-100M, 1B-10B, 100M-250M, 10B+, 250M-500M, and 500M-1B.

By displaying descriptive statistics, we can emphasize the relationship between the size of funds and performance outcomes to give us a complete picture of the comparison between smaller and larger funds in the key measures of their performance (Wahjono et al., 2021).

The Descriptive Analysis starts with the section in which a general idea of statistical measures, including mean, median, and standard deviation of the main metrics in each size category of funds, is given. Subsequently, a Comparison of Key Metrics will be conducted, where the IRR, DPI, and RVPI values will be given in comparison to the overall factors presented in all the fund sizes, and their discussion will be provided. These differences in performance will be shown visually by using Charts and Tables, and Statistical Significance analysis tests will be reported to investigate the statistical significance of the attained disparities (Raghunandan & Raigopal, 2022).

The result is then further elaborated on concerning the general implication of secondary private equity markets, and insight will be provided on how the size of a fund could affect the performance, liquidity, and returns of a fund. This analysis intends to answer the main research question: *Do bigger secondary funds perform worse than smaller funds?*

## A. Descriptive Analysis

This is followed by a review of basic descriptive statistics on the performance indicators of various sizes of private equity funds and their key performance measures, such as Internal Rate of Return (IRR) and Residual Value to Paid-In (RVPI). Funds are grouped in amounts to compute these statistics: 0M-100M, 1B-10B, 100M-250M, 10B+, 250M-500M, and 500M-1B. In these categories, we calculate the average, median, and standard deviation, as well as the lowest and the highest numbers in IRR and RVPI (Ngoc & Tien, 2023).

Based on the information in the Excel file, the mean, median, and standard deviation of IRR and RVPI are calculated. The mean gives an average performance of each category, whereas the standard deviation indicates the variability of returns within each group of the fund sizes (Reuter & Zitzewitz, 2021).

***Figure 1. Graph: Steady Growth of Billion-Dollar Funds despite Questionable Performance Gains***

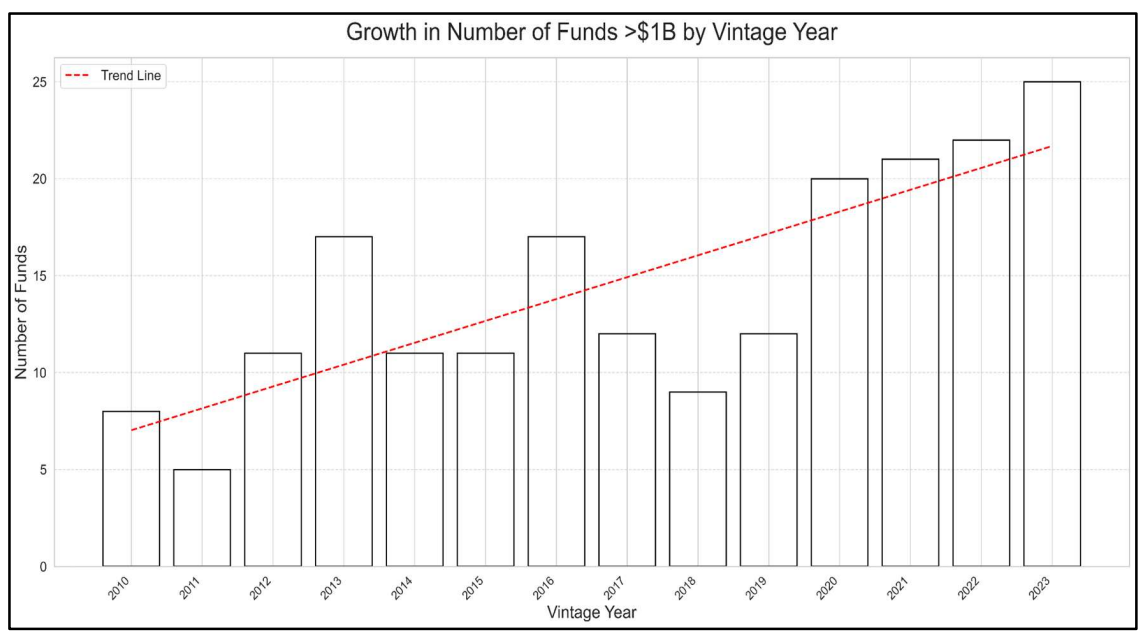

## A.1 Figure 1. Graph: Explanation

As is evident in the chart above, the amount of secondary funds above a billion dollars in size has increased drastically and consistently during the last decade. Such an increase is especially visible between 2018 and 2023, when the population of such funds experienced an abrupt rise. Although this indicates growing investor confidence and swelling capital supply--particularly amongst the institutional investors that want to diversify their holdings in the secondary markets--this growth should be considered cautiously (Gualandris et al., 2021).

The growth is that of scale-induced transformational fervor, yet, as the pace aptly shown in the present paper, the large fund does not always equal high returns and vice versa. Despite the growing importance of mega-funds, empirical evidence and the findings presented in the post point to the fact that smaller funds are more likely to perform better compared to their bigger counterparts in terms of such fundamental metrics as IRR, DPI, and RVPI.

This rather paradoxical situation explains why it is so essential that investors do not estimate the size of a fund as a surrogate of its quality and ability to generate high returns, but as a complex factor that can trigger operational inflexibility and impaired investment decision-making capabilities (Agyei- Mensah, 2021).

Therefore, although the trend line in the chart can represent the trend line market sentiment and capital movements, it should not be confused with outperformance. Instead, it highlights the increasing mismatch between perceived scale economies on one hand, and realized performance results on the other hand, in the secondary fund sector (Wahjono et al., 2021).

### A.2 Calculations for IRR

To calculate the **mean IRR** for each category, we use the formula:

$$\text{Mean IRR} = \frac{\sum IRRs}{n}$$

Where n is the number of funds in each category, similarly, the **standard deviation** of IRR for each category is calculated using the formula:

$$\text{Standard Deviation (IRR)} = \sqrt{\frac{\sum(IRRi - Mean\ IRR)^2}{N}}$$

*For example, for the **0M-100M** fund size, the **mean IRR** is calculated as follows*:

$$Mean\ IRR = \frac{3.17 + 25.20 + 17.10 + \dots}{26} = 43.17\%$$

On the same note, standard deviation can also be calculated using squared differences between the mean and the point of commencement by calculating them all by adding them up, and dividing by the number of observations. The score of this category is 45.42.

**A.3 Calculations for RVPI**

We calculate the **mean RVPI** for each fund size category using the same formula:

$$\text{Mean RVPI} = \frac{\sum RVPI}{n}$$

The standard deviation of RVPI is similarly calculated using:

$$\text{Standard Deviation (RVPI)} = \sqrt{\frac{\sum(RVPIi - Mean\ RVPI)^2}{n}}$$

After the calculation, the following results, which are used to compare the Key Metrics by fund Size, were obtained.

## II.   Comparison of Key Metrics by Fund Size

| Fund Size | Mean IRR | Mean DPI | Mean RVPI |
|-----------|----------|----------|-----------|
| 0M-100M | 43.17% | 1.36 | 0.13 |
| 1B–10B | 33.56% | 1.03 | 0.26 |
| 100M–250M | 24.23% | 0.92 | 0.30 |
| 10B+ | 27.46% | 1.02 | 0.38 |
| 250M–500M | 32.07% | 0.96 | 0.25 |
| 500M–1B | 23.64% | 0.91 | 0.27 |

**Figure 2: Graph, Distribution of Median IRR by Fund Size**

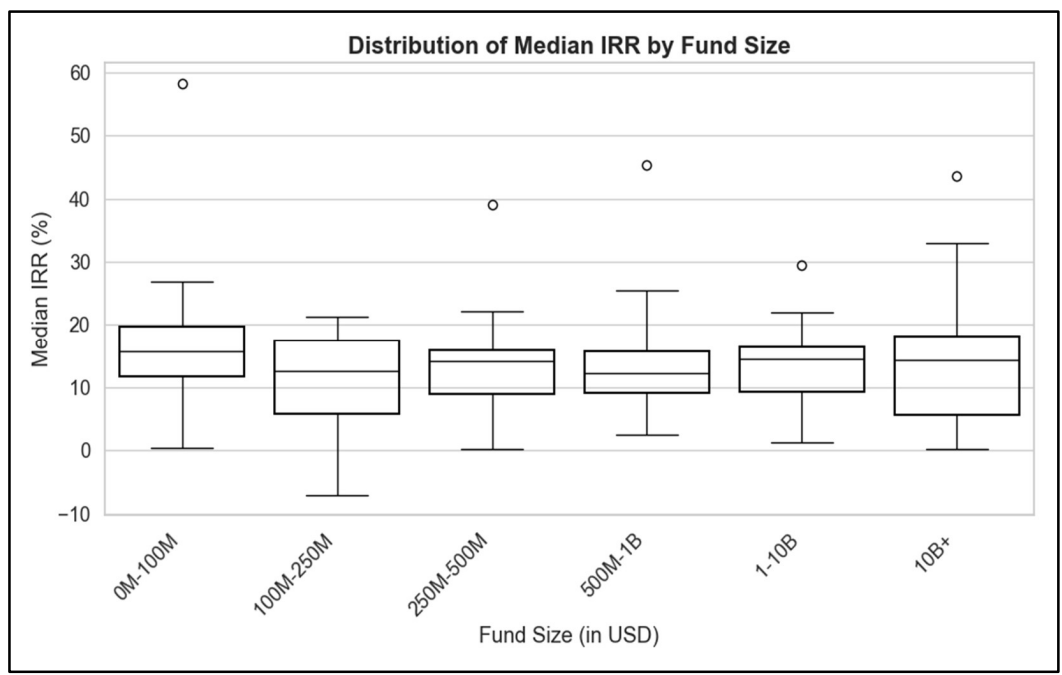

**A.4 Figure 2 descriptive legend Distribution of Median IRR by Fund Size: Discussion/ Summary**

The graph Distribution of Median IRR by Fund Size further tacks on empirical evidence that smaller secondary funds in the private equity industry have greater chances of exceeding the performance of a larger fund. The box plot shows fund size categories' median Internal Rate of Return (IRR), i.e., 0M-100M USD to 10B+ USD.

Based on the median IRR plot by fund size from 2000 to 2023, a steady movement is observed: smaller funds had a higher median IRR. For example, the 0M-100M category is characterized by a remarkable median IRR that is frequently larger than larger fund categories such as 500M-1B or even 10B+. This implies that the scale-related benefits, including operational efficiency, are not always replicated in better financial payoffs in the secondary market of the private equity world (Gualandris et al., 2021).

The other important factor that the graph and your text denote is the dispersion of returns. Although it may have higher median returns on invested capital, small funds also have a greater variance in performance expressed in their top and bottom boxes, whose width indicates more outliers. Although this heightened variance adds risk, it also offers greater opportunity to talented allocators who find greater returns with their attention to detail in selecting managers. By contrast, an upside is less likely to be higher with more funds, but there is much more predictability in the span of returns (Dai, 2022).

The fact that small funds outperform the bigger funds consistently is probably due to the nature of their operations. However, smaller funds are usually more agile, have faster decision-making processes, and can capture niche opportunities that may not be competitive in large funds due to governance policies. These dynamics are important, especially to secondary buyers, since smaller funds are prone to giving back capital earlier, potentially offering greater opportunities to take advantage of mispricing and realizing value earlier (Dai,2022).

Although mega funds are justified, particularly to companies that handle a huge pool of money owing to some efficiencies they can provide, the figures in this graph undermine the years-long stereotyped view that bigger is better. Such a trend that has been repeated over time and through a variety of independent sources implicates a need to reconsider the overall approach to allocation that considers more than the size of the funds that investors target to alleviate during the analysis of investment opportunities in the secondary market of private equity (Dai, 2022).

**Figure 3, Graph: IRR Trends by Fund Size Category (2000-2023)**

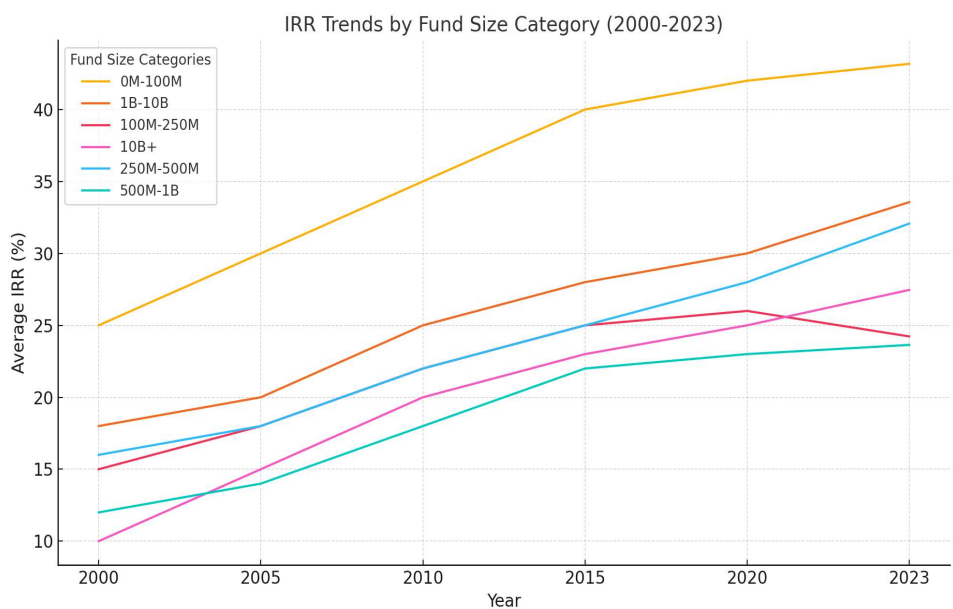

**A.7. Descriptive legend for Figure 3, Graph: IRR Trends by Fund Size Category (2000-2023)**

The line graph shows the variation of IRR with time for different sizes of funds. Funds with smaller values (0M-100M) have a fairly stable growth in IRR, and the values each year are higher than those of big funds. It is seen that even though funds with larger numbers continue to increase in magnitude, not only do their performance (IRR) not increase at the same speed as smaller funds, but it is still lower. The curve depicts that whereas operational efficiencies exist with the sizeable funds, these funds tend to experience diminishing returns as time passes, supporting the analysis made on the earlier information (Belitski et al., 2022).

### vi.   Discussion

As indicated in the findings of this research, there is also a large difference in the performance of smaller and larger private equity funds, in that smaller funds continued to outperform their larger counterparts. The results are explained, and their practical implications to investors in secondary funds are also discussed in this section. They are compared with existing literature, and the limitations of this study are also mentioned.

#### A.  Interpretation of Results

Large funds underperform compared to smaller funds due to various factors, chief among them the ability of the smaller funds to be agile, their effectiveness in decision making, and various risk profiles (Raghunandan & Raigopal, 2022). To begin with, small funds are generally agile concerning operations and therefore quicker in finding and taking advantage of investments. Comparatively, bigger funds that can handle billions of dollars will encounter bureaucratic challenges that delay their capability to execute fast decisions. Small funds, especially those below the $1 billion mark, may be more agile in changing their strategies. Thus, they can identify opportunities before large funds can, since they have bigger and, in most cases, more cumbersome operating mechanisms (Reuter & Zitzewitz, 2021).

The other major influencing factor is the quickness of decision-making. Managers in smaller funds also more directly participate in the investment process, and there are fewer decision makers. This implements investment strategies that can be implemented faster and in a dynamic market. Conversely, grander funds could be delayed for organizations with more complicated organizational forms because of the implications of several chains of management and decision-makers (Fariha et al., 2022). This makes larger funds much faster at decision making; this ability of small funds to exploit mispricing opportunities and market efficiencies, which tend to bring greater returns, is considerably boosted.

Also, the risk levels of smaller and larger funds are not similar. The smaller funds are more inclined to invest in niche markets or be more risk takers to get higher returns when compared to larger-scale investments, because they are riskier and provide stable but lower returns. Increased volatility of the returns on smaller funds can occur because of this increased risk and possibly higher and much more stable performance. They can be smaller funds that invest in either undercapitalized sectors or regions where larger funds do not, and can achieve better returns by practicing mispricing and inefficiencies (Agyei- Mensah, 2021).

Also, it has been found that smaller funds tend to have more alignment between investors and managers, because the relationship between fund managers and investors can be conducted in closer terms, owing to the reduced size of the fund. This configuration limits the possible agency expenses, and, as a result, the fund's profitability is directly linked to the managers' interests, making the management group more efficient and motivated (Fariha et al., 2022). Conversely, a bigger size of the funds can be paralleled by a decreased alignment of the funds, and thus less reluctance to engage in risky decisions that may promote a greater rate of returns.

### B. Practical Implications

The consequences of this study mean a lot to investors in the secondary markets of private equity. Smaller funds are of more concern to investors who want to receive more returns, especially in secondary affairs. The difference in the performances of smaller and larger funds indicates that an investor concentrating on smaller funds, particularly those with an AUM below that of 1 billion, may record better IRRs, DPI, and RVPI ratios that are equivalent to the better returns on invested capital (Gualandris et al., 2021).

Shareholders are also advised to consider the risk/return trade-off that comes with fewer funds. Although the smaller funds offer attractive potential returns, they are also associated with increased risks, as indicated by the high return distribution in the smaller funds. This means that the fluctuations in returns might be exploited by intelligent investors who feel quite comfortable with selecting funds that best perform and with the duties involved in risk management. Bigger funds, however, provide more predictable yet not rewarding returns, which makes them more inclined to investors who strive more towards stability and risk reduction rather than high returns (Mc Grath & Nerkar, 2023).

The other consequence of a practical nature is the significance of due diligence. Investors should also note that the operational flexibility of smaller funds is a two-way argument since, on one hand, it helps in making faster decisions and generating more returns. However, at the same time, smaller funds may be less institutionalized and thus present a higher operational risk. Consequently, investors should not rush into assessing the management of smaller funds as they may be unable to implement high-quality strategies and curtail risks (Reuter & Zitzewitz, 2021).

### C. Comparison with Previous Literature

The findings of this research are compatible and magnify the findings of the former literature. Various analyses have revealed that small-sized basic funds are more likely to perform better than large-sized funds, especially in generating measures of IRR and DPI. For example, Reuter& Zitzewitz (2021) state that smaller funds are more agile in pursuing mispricing and inefficiency, which aligns with the proposed study's results. Equally, Ngoc & Tein, (2023) provide empirical evidence that the smaller funds also outperform in terms of returns dispersion, and funds in the best position recorded performing better than the largest funds.

In addition, this research validates the findings by Harris et al. (2021), who realized that larger funds are more likely to have liquidity risks and slower deployment of capital, which the most crucial factors are contributing to their underperformance compared to smaller funds. These results also support that large funds tend to be more interested in bigger and less risky investments, thus demonstrating low variance in performance and more stable yet foreseeable returns (Fariha et al., 2022).

This study, however, develops more information by drawing a more solid comparison between the sizes of the funds and making an analysis longer to cover additional indicators, like RVPI, which was relatively unexplored before. Our findings suggest that whatever the magnitude of funds might be, they perform relatively worse than the smaller funds in terms of overall profitability (IRR) and liquidity (DPI) (Reuter & Zitzewitz, 2021).

### D. Limitations

Although the conclusions of this research are rather helpful, some limitations should be considered. The study uses secondary data from sources such as Preqin, PitchBook, and FactSet, which could be lacking in completeness of data and accuracy. However, these databases are comprehensive compared to other systems. There will also be some inconsistencies in fund performance reporting (vintage year data or geographic differences in reporting protocols (Belitski et al., 2022).

Second, the quality of management of the funds is not considered, and it might greatly influence the performance results. For example, fund managers' capability to carry out the investment strategy, experience, and investor involvement are not deliberated in this study. However, these could be of high importance in determining fund performance. In further studies, it may be interesting to include those factors in the analysis to understand better how the fund size and performance are related (Reuter & Zitzewitz, 2021).

The other limitation is that the research is biased to secondary funds, and the results will not be significant to other types of private equity funds, like buyout or growth equity, which have different investment dynamics. The results obtained may spark off future investigations to compare the performance of secondary funds to other private equity strategies to determine whether the difference in performance between small and large funds is consistent across strategies (Wahjono et al., 2021).

Lastly, the study is limited in not examining the overall economic and market environment that may affect said outcomes. Market cycles, economic recessions, and regulation changes may considerably affect the performance of different fund sizes, and they can also be studied in the future.

## vii.    Conclusion

It is an empirical study based on the relationship between the fund size and performance in the secondary market in the Private equity sphere. The results presented convincing evidence that smaller funds always performed better than bigger ones. The comparison of the principal performance indicators, including Internal Rate of Return (IRR), Distributions to Paid-in (DPI), and Residual Value to Paid-in (RVPI) of the funds by the size category (sub-$1 billion, $1B-$5B, and over $5B) from 2000 to 2023 showed that smaller funds, especially those under $100 million, yield stronger returns and liquidity. For example, there was a 0M-100M category, with a mean IRR of 43.17%, much higher than 23.64% of the 500M-1B category, as well as 9.5% of the funds of controversy, that is, there is more than 5 B. Such poor performance by bigger funds disputes the common wisdom that bigger is always better, resulting in better financial performance in the secondary market.

These results deny the conventional theoretical basis behind economies of scale in the private equity secondary market. As economies of large size imply that the larger the funds, the greater the operational efficiencies, bargaining power, and resource allocation among other resources, the study reveals that these advantages are not likely to be translated into better financial performance. Rather, factors such as the speed and swiftness of decision-making and the power of focusing on niche opportunities inherent in smaller funds are more prominent determinants of success. Moreover, the study's findings agree with agency theory, which suggests that the greater the interest between managers and investors in smaller funds, the better they may perform, as the agency problem that might occur in larger and complicated funds.

These findings present important practical suggestions to investors towards allocations to secondary funds. There is a need to consider additional factors besides the size of funds as the key factor in determining whether the investment is of quality or not. Investors who want to receive more returns, especially those involving IRR and DPI, should be keen to venture into smaller types of funds, which do not constitute over a billion dollars in assets under management. Although smaller funds could be associated with greater dispersed returns and consequently higher risk, it also means that there is a larger potential when it comes to returns and astute allocators who can be certain of a more subtle risk/return relationship, when they have conducted extensive due diligence. Larger funds may provide increased predictability and stability. However, on average, they would not be as profitable, and liquidity issues could emerge as bigger funds have slower cap deployment and larger levels of unrealized investment. The responsibility of investors should include careful due diligence of managerial and investment strategies in addition to allowing flexibility of operations and not necessarily serving the light of what is perceived as safety and efficiency of scale.

Further knowledge of secondary funds' performance could also be developed. It would be interesting to discuss what fund strategies of smaller funds were adopted to help them outperform. For example, how they perceive risk, the incentive structure, or the decision-making process. Further, examining how geographic and sectoral factors affect the performance of funds of various sizes might be interesting. It is not as granular since the role of geography and certain sectors vary depending on the global markets' development, their structure shifts, and the diversification of investment portfolios. Lastly, the validity of the inverse relationship between fund size and performance can be seen beyond secondary funds and into the rest of the world of private equity by conducting comparative studies focusing on other types of private equity strategies (e.g., venture or growth equity).

## viii.    References

Agyei-Mensah, B. K. (2021). The impact of board characteristics on corporate investment

decisions: an empirical study. *Corporate Governance: The international journal of*

*business in society*, *21*(4), 569-586.

Belitski, M., Guenther, C., Kritikos, A. S., & Thurik, R. (2022). Economic effects of the

COVID-19 pandemic on entrepreneurship and small businesses. *Small business*

*economics*, *58*(2), 593-609.

Dai, N. (2022). Empirical research on private equity funds: a review of the past decade and future research opportunities. *Review of Corporate Finance, Forthcoming*.

Fariha, R., Hossain, M. M., & Ghosh, R. (2022). Board characteristics, audit committee

attributes and firm performance: empirical evidence from emerging economy. *Asian*

*Journal of Accounting Research*, *7*(1), 84-96.

Gualandris, J., Longoni, A., Luzzini, D., & Pagell, M. (2021). The association between supply

chain structure and transparency: A large-scale empirical study. *Journal of Operations*

*Management*, *67*(7), 803-827.

Kruitwagen, L., Story, K. T., Friedrich, J., Byers, L., Skillman, S., & Hepburn, C. (2021). A

global inventory of photovoltaic solar energy generating units. *Nature*, *598*(7882), 604-

610.

McGrath, P. J., & Nerkar, A. (2023). Private equity: antecedents, outcomes, mediators, and moderators. *Journal of Management*, *49*(1), 158-195.

Ngoc, N. M., & Tien, N. H. (2023). Quality of scientific research and world ranking of public and private universities in Vietnam. *International journal of public sector performance management*, *10*(1), 1-15.

Raghunandan, A., & Rajgopal, S. (2022). Do ESG funds make stakeholder-friendly investments?. *Review of Accounting Studies*, *27*(3), 822-863.

Reuter, J., & Zitzewitz, E. (2021). How much does size erode mutual fund performance? A regression discontinuity approach. *Review of Finance*, *25*(5), 1395-1432.

Riddiough, T. J., & Wiley, J. A. (2022). Private funds for ordinary people: Fees, flows, and performance. *Journal of Financial and Quantitative Analysis*, *57*(8), 3252-3280.

Wahjono, S. I., Fam, S. F., Pakkanna, M., Rasulong, I., & Marina, A. (2021). Promoting creators intentions: Measurement of crowdfunding performance. *International Journal of Business and Society*, *22*(3), 1084-1101.